\documentclass[aps,prc,superscriptaddress,10pt,showpacs,notitlepage,  
nofootinbib
]{revtex4-1}
\usepackage{bm}
\usepackage{amsmath,amssymb,amsthm}
\usepackage{latexsym,graphicx,color,subfigure}
\usepackage{enumerate}
\usepackage{amssymb}
\usepackage{hyperref}
\usepackage{mathtools}

\usepackage{graphicx}
\graphicspath{ {images/} }

\usepackage{color}

\newcommand{\be}{\begin{equation}}
\newcommand{\ee}{\end{equation}} 
\newcommand{\beq}{\begin{eqnarray}}
\newcommand{\eeq}{\end{eqnarray}}

\newcommand{\bea}{\begin{eqnarray}}
\newcommand{\eea}{\end{eqnarray}}

\def\tc{\textcolor{red}}
\def\tc2{\textcolor{blue}}
 
\renewcommand{\vec}[1]{\boldsymbol{#1}}

\newcommand{\dtr}{\mathrm{tr}}

\allowdisplaybreaks[3]
\def\simge{\mathrel{
   \rlap{\raise 0.511ex \hbox{$>$}}{\lower 0.511ex \hbox{$\sim$}}}}
\def\simle{\mathrel{
   \rlap{\raise 0.511ex \hbox{$<$}}{\lower 0.511ex \hbox{$\sim$}}}}
\def\bigs{\mathrel{
   \rlap{\raise 0.531ex \hbox{$>$}}{\lower 0.531ex \hbox{$<$}}}}

\usepackage[normalem]{ulem}

\renewcommand\sout{\bgroup \color{blue} \ULdepth=-.5ex \ULset}

\begin{document}
\title{Coexistence phase of $^{1}S_{0}$ and $^{3}P_{2}$ superfluids in neutron stars}
\author{Shigehiro Yasui}
\email{yasuis@keio.jp}
\affiliation{Research and Education Center for Natural Sciences, Keio University, Hiyoshi 4-1-1, Yokohama, Kanagawa 223-8521, Japan}
\author{Daisuke Inotani}
\email{dinotani@keio.jp}
\affiliation{Research and Education Center for Natural Sciences, Keio University, Hiyoshi 4-1-1, Yokohama, Kanagawa 223-8521, Japan}
\author{Muneto Nitta}
\email{nitta(at)phys-h.keio.ac.jp}
\affiliation{Research and Education Center for Natural Sciences, Keio University, Hiyoshi 4-1-1, Yokohama, Kanagawa 223-8521, Japan}
\affiliation{Department of Physics, Keio University, Hiyoshi 4-1-1, Yokohama, Kanagawa 223-8521, Japan}
\date{\today}
\begin{abstract}
In neutron star matter, there exist $^{1}S_{0}$ superfluids in lower density in the crust while $^{3}P_{2}$ superfluids are believed to exist at higher density deep inside the core. In the latter, depending on the temperature and magnetic field, either the uniaxial nematic phase, the D$_{2}$-biaxial nematic phase, or the D$_{4}$-biaxial nematic phase appears. In this paper, we discuss a mixture of the $^{1}S_{0}$ and $^{3}P_{2}$ superfluids and find their coexistence.  Adopting the loop expansion and the weak-coupling approximation for the interaction between two neutrons, we obtain the Ginzburg-Landau (GL) free energy in which both of the $^{1}S_{0}$ and $^{3}P_{2}$ condensates are taken into account by including the coupling terms between them. We analyze the GL free energy and obtain the phase diagram for the temperature and magnetic field. We find that the $^{1}S_{0}$ superfluid excludes the $^{3}P_{2}$ superfluid completely in the absence of magnetic field, they can coexist for weak magnetic fields, and the $^{1}S_{0}$ superfluid is expelled by the $^{3}P_{2}$ superfluid at strong magnetic fields, thereby proving the robustness of  $^{3}P_{2}$ superfluid against the magnetic field. We further show that the D$_{4}$-BN phase covers the whole region of the $^{3}P_{2}$ superfluidity as a result of the coupling term, in contrast to the case of a pure  $^{3}P_{2}$ superfluid studied before in which the D$_{4}$-BN phase is realized only under strong magnetic fields. Thus, the D$_{4}$-BN phase is topologically the most interesting phase, e.g., admitting half-quantized non-Abelian vortices relevant not only in magnetars but also in ordinary neutron stars. 
\end{abstract}
\maketitle

\section{Introduction}
\label{sec:introduction}

Neutron stars are compact stars under extraordinary conditions, 
providing astrophysical laboratories to study exotic phases of nuclear matter at high density, under rapid rotation and with a strong magnetic field 
(see Refs.~\cite{Graber:2016imq,Baym:2017whm} for recent reviews). 
Some advanced results in the recent astrophysical researches are given by
the reports on the observations of highly massive neutron stars the masses of which reach almost twice as large as the solar mass~\cite{Demorest:2010bx,Antoniadis1233232} and
the detections of the gravitational waves from a binary neutron star merger~\cite{TheLIGOScientific:2017qsa}.
Inside neutron stars, one of the most important ingredients is neutron superfluidity and proton superconductivity (see Refs.~\cite{Chamel2017,Haskell:2017lkl,Sedrakian:2018ydt} for recent reviews).
The superfluid and superconducting components can alter excitation modes, affecting several properties in neutron stars,  e.g., neutrino emissivities and specific heats relevant, respectively, to the thermal evolution of  neutron stars and to a long relaxation time after sudden speed-up events (glitches) of neutron stars~\cite{Baym1969,Pines1972,Takatsuka:1988kx}.
For example, the enhancement of neutrino emissivity near the critical point of the superfluid transition was studied in detail~\cite{Yakovlev:2000jp,Potekhin:2015qsa,Yakovlev:1999sk,Heinke:2010cr,Shternin2011,Page:2010aw}.
It was also proposed that glitches of pulsars may be explained by quantized vortices in superfluids~\cite{reichley,Anderson:1975zze}.

The neutron superfluids are induced by the attractive forces between two neutrons in several different channels in the low-density regime around the crust~\cite{Chamel:2008ca} (see also Ref.~\cite{Dean:2002zx} and the references therein).
The superfluids are often studied by a mean-field theory in the weak-coupling limit.
It should be noted, however, that
fluctuations become important for neutron $^{1}S_{0}$ superfluidity, 
as the case of the Bardeen-Cooper-Schrieffer (BCS)-Bose-Einstein condensation (BEC) crossover phenomenon in ultracold atomic Fermi gases (see Ref.~\cite{Strinati:2018wdg} and the references therein).
Recently, the fluctuation effects on the neutron $^{1}S_{0}$ pairing were studied in the framework of the Nozi{\`e}res and Schmitt-Rink scheme~\cite{Nozieres1985}.\footnote{See, e.g., Refs.~\cite{RevModPhys.82.1225,Mueller_2017,Jensen2019} for the applications of the BCS-BEC crossover phenomena in cold atom physics.}
In this scheme, the effects of pairing fluctuations were studied for the equation of state in neutron matter by considering the finite effective range as well as strong-coupling effects~\cite{PhysRevA.97.013601}.
In Ref.~\cite{Tajima:2019saw}, the pairing fluctuation in the normal state above the critical temperature was studied by adopting the separable potential.
More recently, in Ref.~\cite{Inotani:2019tlo}, the full gap equation was analyzed in detail to obtain the gap strength in neutron $^{1}S_{0}$ superfluids covering a wider range of the density and temperature from zero to the critical temperature.
It was also shown there that the collective modes, i.e., Anderson-Bogoliubov (phase, sound, or phonon) and Higgs (or amplitude) modes, play a remarkable role not only near the critical temperature but also at zero temperature.

Although the $^1S_0$ channel is the most dominant attraction in the low-density regime,
it becomes repulsive in the high-density regime inside the neutron star core.\footnote{Historically, the $^1S_0$ superfluidity at low density was proposed in Ref.~\cite{Migdal:1960}. Later, it was pointed out in Ref.~\cite{1966ApJ...145..834W} that this channel turns out to be repulsive because of the strong repulsion in short range at higher densities.}
In such higher densities, the attraction is provided by the spin-orbit (LS) force in which the angular momentum and the sum of two neutron spins are coupled to each other, and neutron pairs with the total angular momentum $J=2$ with spin triplet and $P$ wave are energetically favored.
The LS force induces the neutron $^{3}P_{2}$ superfluids as the relevant phase in the ground state~\cite{Tabakin:1968zz,Hoffberg:1970vqj,Tamagaki1970,Hoffberg:1970vqj,Takatsuka1971,Takatsuka1972,Fujita1972,Richardson:1972xn,Amundsen:1984qc,Takatsuka:1992ga,Baldo:1992kzz,Elgaroy:1996hp,Khodel:1998hn,Baldo:1998ca,Khodel:2000qw,Zverev:2003ak,Maurizio:2014qsa,Bogner:2009bt,Srinivas:2016kir}.
It is interesting that the neutron $^3P_2$ superfluids can survive under strong magnetic fields, such as in the magnetars with the magnetic field $10^{15}$-$10^{18}$ G.
This is intuitively understood because the spin-$\uparrow\uparrow$ or -$\downarrow\downarrow$ pairs in the spin-triplet pairing cannot be broken by the Zeeman effects.\footnote{The origin of the strong magnetic fields in neutrons stars or in magnetars is still an open problem although there are many theoretical works: spin-dependent interactions~\cite{Brownell1969,RICE1969637,Silverstein:1969zz,Haensel:1996ss}, pion domain walls~\cite{Eto:2012qd,Hashimoto:2014sha}, spin polarizations in quark matter in the neutron star core~\cite{Tatsumi:1999ab,Nakano:2003rd,Ohnishi:2006hs}, and so on. It may be worthwhile to mention that a negative result for the generation of strong magnetic fields was recently announced in a study in terms of the nuclear many-body calculations~\cite{Bordbar:2008zz}.
}
Up to now, the possible existence of neutron $^{3}P_{2}$ superfluids inside the neutron stars has been studied in astrophysical observations.
Recently, it has been pointed out that the neutron $^{3}P_{2}$ superfluids can be relevant for the rapid cooling of a neutron star in Cassiopeia A~\cite{Heinke:2010cr,Shternin2011,Page:2010aw}.
In fact, the enhancement of neutrino emissivities can be caused by the formation and dissociation of neutron $^3P_2$ Cooper pairs.
The neutron $^3P_2$ superfluids are interesting also in terms of the condensed-matter physics.
Several theoretical studies show that the neutron $^3P_2$ superfluids have rich structures in the condensates because of a variety of combinations of spin-triplet and $P$-wave angular momentum in the Cooper pairs. 
The superfluid states with $J=2$ are classified into nematic, cyclic, and ferromagnetic phases~\cite{Mermin:1974zz}.
Among them, the nematic phase is the ground state in the weak-coupling limit of $^3P_2$ superfluids~\cite{Fujita1972,Richardson:1972xn,Sauls:1978lna,Muzikar:1980as,Sauls:1982ie,Vulovic:1984kc,Masuda:2015jka,Masuda:2016vak}.
The nematic phase consists of the three subphases:
the uniaxial nematic (UN) with the U(1) symmetry, and dihedral-two and dihedral-four biaxial nematic (D$_{2}$-BN and D$_{4}$-BN) phases with  the D$_{2}$ and D$_{4}$ symmetries, respectively.\footnote{See, e.g., Appendix~B in Refs.~\cite{Yasui:2019unp,Yasui:2019pgb} for detailed information on the UN, D$_{2}$-BN, and D$_{4}$-BN phases.}

The $^3P_2$ superfluids allow bosonic excitations as collective modes~\cite{Bedaque:2003wj,Leinson:2011wf,Leinson:2012pn,Leinson:2013si,Bedaque:2012bs,Bedaque:2013rya,Bedaque:2013fja,Bedaque:2014zta,Leinson:2009nu,Leinson:2010yf,Leinson:2010pk,Leinson:2010ru,Leinson:2011jr},
which are considered to be relevant to the cooling process by neutrino emissions from neutron stars.\footnote{Note that the cooling process is related not only to low-energy excitation modes but also to quantum vortices~\cite{Shahabasyan:2011zz}.}
Bosonic excitations can be best discussed by
the Ginzburg-Landau (GL) theory as a bosonic effective theory around the transition point from the normal phase to the superfluid phase~\cite{Fujita1972,Richardson:1972xn,Sauls:1978lna,Muzikar:1980as,Sauls:1982ie,Vulovic:1984kc,Masuda:2015jka,Chatterjee:2016gpm,Masuda:2016vak,Yasui:2018tcr,Yasui:2019tgc,Yasui:2019unp,Yasui:2019pgb,Yasui:2019vci,Mizushima:2019spl}.
This is regarded as the low energy effective theory of the Bogoliubov--de Gennes (BdG) equation, which is a fundamental theory in terms of  fermions~\cite{Tabakin:1968zz,Hoffberg:1970vqj,Tamagaki1970,Takatsuka1971,Takatsuka:1992ga,Amundsen:1984qc,Baldo:1992kzz,Elgaroy:1996hp,Khodel:1998hn,Baldo:1998ca,Khodel:2000qw,Zverev:2003ak,Maurizio:2014qsa,Bogner:2009bt,Srinivas:2016kir,Mizushima:2016fbn,Mizushima:2019spl,Masaki:2019rsz}.
The GL equation can be obtained by a systematic expansion of  the fermion loops by integrating out the fermionic degrees of freedom.
The GL equation is expressed by a series of the power terms of the order parameter in the $^{3}P_{2}$ superfluids.
Usually the GL expansion up to the fourth order is enough to determine 
the ground state. 
However, in the case of  $^{3}P_{2}$ superfluids,
the expansion up to the fourth order cannot determine the ground state uniquely, 
due to a continuous degeneracy among 
the UN, D$_{2}$-BN, and D$_{4}$-BN phases.\footnote{At the fourth order, in fact, an SO(5) symmetry happens to exist in the potential term as an extended symmetry absent in the original Hamiltonian. It is known that, in this case, the spontaneous breaking of such extended symmetry eventually generates a quasi-Nambu-Goldstone mode 
\cite{Uchino:2010pf}.
}
This degeneracy can be resolved by including the sixth-order terms in the GL expansion, and the ground state is determined uniquely~\cite{Masuda:2015jka}.
However, the sixth-order term 
brings the instability for a large value of the order parameter,
and the system becomes unbounded and unstable.
This problem can be cured if the eighth-order terms are included 
\cite{Yasui:2019unp}.
Therefore, the expansion up to the eighth order is
the first to allow the globally stable unique ground state. 
As a by-product of this expansion, 
the (tri)critical end point (CEP) separating 
the first- and second-order phase transition between 
the D$_2$ and D$_4$ BN phases 
was found in the phase diagram~\cite{Mizushima:2019spl}, 
which was known  before to exist in the BdG equation~\cite{Mizushima:2016fbn}.

The GL equation is easily applied to describe nonuniform condensations.
In Ref.~\cite{Yasui:2019vci}, the GL equation was adopted to investigate the position dependence of the order parameter in the neutron $^{3}P_{2}$ superfluids (the quasistable domain walls).
The GL equation was often used to study 
spontaneously magnetized vortices~\cite{Muzikar:1980as,Sauls:1982ie,Fujita1972,Masuda:2015jka},
solitonic excitations on a vortex~\cite{Chatterjee:2016gpm}, half-quantized non-Abelian vortices~\cite{Masuda:2016vak}, and
topological defects (boojums) on the boundary of $^3P_2$ superfluids~\cite{Yasui:2019pgb}. 
The topological properties in neutron $^{3}P_{2}$ superfluids share common interests in condensed-matter physics: $D$-wave superconductors~\cite{Mermin:1974zz}; $P$-wave superfluidity in $^{3}$He liquid~\cite{vollhardt2013superfluid,volovik}; chiral $P$-wave superconductivity, e.g., in Sr$_2$RuO$_4$~\cite{RevModPhys.75.657}; and spin-2 Bose-Einstein condensates~\cite{2010arXiv1001.2072K}. 
The boojums on the boundary of $^3P_2$ superfluids 
\cite{Yasui:2019pgb}  
share some properties with similar objects on the boundary of  spin-2 Bose-Einstein condensations~\cite{2019arXiv190702216C}
and  liquid crystals~\cite{Urbanski_2017}. 

As presented so far, the $^{1}S_{0}$ superfluids are dominated at the lower-density regime and the $^{3}P_{2}$ superfluids are dominated at the higher-density regime, because the interaction strength changes according to the scattering energy of two neutrons.
Due to the fact that the $^{3}P_{2}$ channel is always attractive from the low density to the high density, we should reasonably expect that there can be the intermediate-density regime in which the attraction in the $^{3}P_{2}$ channel becomes comparable to the attraction in the $^{1}S_{0}$ channel before the $^{1}S_{0}$ channel becomes repulsive.
In such a density region, we can consider the situation that the $^{1}S_{0}$ and $^{3}P_{2}$ superfluids coexist for given temperature and magnetic field.
In the context of the condensed-matter physics, the mixture of $S$-wave and $P$-wave condensates is discussed in non-centrosymmetric superconductors~\cite{bauer2012non} such as CePt$_{3}$Si~\cite{PhysRevLett.92.027003} and ultracold Fermi gases with a spin-orbit coupling~\cite{PhysRevLett.107.195304}.

The purpose of the present paper is to reveal the possibility of the coexistence of the $^{1}S_{0}$ and $^{3}P_{2}$ superfluids when the coupling between both the superfluids is taken into account.
 Such information will be important for the study of internal structures of neutron stars.
We first introduce the Lagrangian for the interacting neutrons in the $^{1}S_{0}$ and $^{3}P_{2}$ channels and derive the GL free energy by adopting the loop expansion and the weak-coupling approximation for the neutrons.
We then analyze the GL free energy and show the phase diagram that the $^{1}S_{0}$ superfluid completely excludes the $^{3}P_{2}$ superfluid with zero magnetic field, both the phases can coexist in the weak-magnetic field region, and the $^{1}S_{0}$ superfluid is expelled by the $^{3}P_{2}$ superfluid in the strong-magnetic field region.
We also find that the D$_{4}$-BN phase covers the whole region 
of the $^{3}P_{2}$ superfluidity 
as a result of the coupling term, 
in contrast to the case of a pure  $^{3}P_{2}$ superfluid in which 
the D$_{4}$-BN phase is realized only under strong magnetic fields.
These results can be understood from the fact that $\uparrow\downarrow$ pairs exist and are broken by the Zeeman effects 
for the $^1S_0$ superfluid, 
and the UN and D$_2$-BN phases in the $^{3}P_{2}$ superfluid, while 
all the neutrons form $\uparrow\uparrow$ or $\downarrow\downarrow$ pairs, which are robust against the magnetic field, with equal fraction 
in the D$_{4}$-BN phase.
We thus can expect that the D$_{4}$-BN phase is relevant not only in magnetars but also 
in ordinary neutron stars. This is phenomenologically interesting because 
the D$_{4}$-BN phase is topologically rich, admitting, for instance, 
non-Abelian vortices~\cite{Masuda:2016vak}. 

The paper is organized as follows.
In Sec.~\ref{sec:formalism}, we derive the GL free energy of the mixture of $^{1}S_{0}$ and $^{3}P_{2}$ superfluids.
In Sec.~\ref{sec:numerical_results}, we present the phase diagram 
in terms of the temperature and magnetic field.
The final section is devoted to our conclusion and outlook.

\section{Formalism}
\label{sec:formalism}
In this section, we derive the GL theory for 
the mixture of $^{1}S_{0}$ and $^{3}P_{2}$ superfluids 
by integrating out fermion fields.
To this end, 
for the neutron two-component spinor field $\varphi(t,\vec{x})$, we consider the Lagrangian
\begin{align}
   {\cal L} = {\cal L}_{0} + {\cal L}_{\mathrm{int}},
\end{align}
where the free part of the Lagrangian is given by
\begin{align}
   {\cal L}_{0}
=
   \varphi(t,\vec{x})^{\dag} \biggl( i\partial_{t} + \frac{\vec{\nabla}^{2}}{2m} + \mu \biggr) \varphi(t,\vec{x}),
\end{align}
and the interaction part is given by
\begin{align}
   {\cal L}_{\mathrm{int}}
=
   - V_{\mathrm{Scalar}} - V_{\mathrm{LS}} - V_{B},
\end{align}
for the scalar interaction, the LS interaction, and the magnetic term.
Here $m$ is the neutron mass and $\mu$ is the chemical potential.
In the following, we consider the momentum space, 
in which
the scalar potential term is given by
\begin{align}
   V_{\mathrm{Scalar}}
=
   \varphi_{\vec{q}i}^{\ast} \varphi_{\vec{p}j} \bigl(V_{\mathrm{Scalar}}\bigr)_{ijkl} \varphi_{-\vec{q}k}^{\ast} \varphi_{-\vec{p}l},
\label{eq:scalar_potential_momentum}
\end{align}
where
$
   \bigl(V_{\mathrm{Scalar}}\bigr)_{ijkl}
=
   g_{0} \, \delta_{ij}\delta_{kl} + g_{1} \sigma^{a}_{ij}\sigma^{a}_{kl}
$
with the coupling constants $g_{0}$ and $g_{1}$ for the spin-independent and spin-dependent terms, respectively, $i,j,k,l=\uparrow,\downarrow$ are the spin of a neutron, and $\vec{\sigma}=(\sigma^{1},\sigma^{2},\sigma^{3})$ are the Pauli matrices.
We take the summation for the repeated indices, $a=1,2,3$.
$\vec{p}$ and $\vec{q}$ denote the three-dimensional momenta in the center-of-mass system for the two neutrons.
The LS potential term is given by
\begin{align}
   V_{\mathrm{LS}}
=
   \varphi_{\vec{q}i}^{\ast} \varphi_{\vec{p}j} \bigl(V_{\mathrm{LS}}\bigr)_{ijkl} \varphi_{-\vec{q}k}^{\ast} \varphi_{-\vec{p}l},
\end{align}
where
$
  \bigl(V_{\mathrm{LS}}\bigr)_{ijkl} = -ig_{\mathrm{LS}} (\vec{S})_{ijkl} \!\cdot\! \bigl( \vec{p} \times \vec{q} \bigr)
$
with the coupling constant $g_{\mathrm{LS}}$, the total spin $(\vec{S})_{ijkl}=(\vec{s})_{ij}\delta_{kl}+\delta_{ij}(\vec{s})_{kl}$, and the spin operator $\vec{s}=\vec{\sigma}/2$ for a neutron.
The interaction term between the neutron and a magnetic field ($\vec{B}$) is given by
\begin{align}
   V_{B} = -\varphi^{\dag} \vec{m}_{n} \!\cdot\! \vec{B} \varphi,
\end{align}
with the magnetic moment of the neutron
 $\vec{m}_{n} =  -\gamma_{n}\hbar \vec{\sigma}/2$
with the gyromagnetic ratio $\gamma_{n} = 1.20423637941\times 10^{-13}$ MeV T$^{-1}$.

To study the possibility of the coexistence of the $^1S_0$ and  $^3P_2$ superfluids, we apply the bosonization techniques,  i.e., the Stratonovich-Hubbard transformation, both for the $^1S_0$- and the $^3P_2$- pairing channels. Here we introduce the $^{1}S_{0}$ condensate 
$\sigma = G_{\mathrm{S}} \langle \varphi^{t} i\sigma_{2} \varphi \rangle$ with $G_{\mathrm{S}} = - \bigl( g_{0}+3g_{1} \bigr)/2$, 
and 
the $^3P_2$ condensate
$A = G_{\mathrm{T}} \langle T^{ab} \rangle$ with $G_{\mathrm{T}} = g_{\mathrm{LS}}/2$, $T^{ab} = \bigl( \phi^{ab}+\phi^{ba} \bigr)/2-\delta^{ab}\phi^{cc}/3$, and $\phi^{ab} = -\varphi^{t}\Sigma^{a}(\nabla^{b}_{x}\varphi)$ where $\Sigma^{a} = i\sigma^{a}\sigma^{2}$ and $a,b,c=1,2,3$.
Note that $T^{ab}$ denotes the traceless and symmetric tensor.
$\varphi^{t}$ denotes the transpose of $\varphi$, and $\nabla^{a}_{x}$ is defined by $\nabla^{a}_{x} = \partial / \partial x_{a}$.
The brackets mean the thermal expectation values.
For convenience, we use the dimensionless quantities defined by
\begin{align}
   \tilde{\sigma} = \frac{1}{T} \sigma, \quad 
   \tilde{A} = \frac{p_{F}}{T} A, \quad 
   \tilde{x}_{a} = \frac{mT}{p_{F}} x_{a}, \quad 
   \vec{b} = \frac{\gamma_{n}}{(1+F_{0}^{a})T} \vec{B},
\label{eq:dimensionless_quantities}
\end{align}
for the $^{1}S_{0}$ condensate, the $^{3}P_{2}$ condensate, the length scale, and the magnetic field, respectively,
with the Fermi momentum $p_{\mathrm{F}}$ and the temperature $T$. $F_{0}^{a}$ is the Landau parameter in the Fermi liquid theory 
introduced as a correction by the effect of the Hartree-Fock approximation.
This correction is necessary because the Hartree-Fock approximation is not covered in the present calculation for the particle-particle interaction at the one-loop level.

We adopt the one-loop approximation for the effective potential and perform the quasiclassical approximation in the momentum integrals.
Using the abbreviations $\tilde{\sigma}=\tilde{\sigma}(\tilde{\vec{x}})$ and $\tilde{A}=\tilde{A}(\tilde{\vec{x}})$,
we obtain the GL free energy
\begin{align}
   f[\tilde{\sigma},\tilde{A}] = N(0)T^{2} \tilde{f}[\tilde{\sigma},\tilde{A}],
\label{eq:GL_free_energy}
\end{align}
where $N(0)=m p_{\mathrm{F}}/2\pi^{2}$ is the density of states at the Fermi surface.
The dimensionless GL free energy $\tilde{f}[\tilde{\sigma},\tilde{A}]$ consists of the three terms as
\begin{align}
   \tilde{f}[\tilde{\sigma},\tilde{A}]
= \tilde{f}_{S}[\tilde{\sigma}] + \tilde{f}_{P}[\tilde{A}] + \tilde{f}_{SP}[\tilde{\sigma},\tilde{A}],
\label{eq:free_energy_twice_dimensionless}
\end{align}
where each term is defined by
\begin{align}
   \tilde{f}_{S}[\tilde{\sigma}]
&=
    \tilde{\alpha}_{S}^{(0)}
    \tilde{\sigma}^{\ast} \tilde{\sigma}
 + \tilde{K}_{S}^{(0)}
    \bigl( {\nabla}_{\tilde{x}i}\tilde{\sigma}^{\ast} \bigr) \bigl( {\nabla}_{\tilde{x}i} \tilde{\sigma} \bigr)
 + \tilde{\beta}_{S}^{(0)}
    \tilde{\sigma}^{\ast 2} \tilde{\sigma}^{2}
 + \tilde{\beta}_{S}^{(2)}
    |\vec{b}|^{2} \tilde{\sigma}^{\ast} \tilde{\sigma},
\label{eq:free_energy_twice_dimensionless_S}
\end{align}
for the $^{1}S_{0}$ condensate,
\begin{align}
 \tilde{f}_{P}[\tilde{A}]
&=
   \tilde{\alpha}_{P}^{(0)}
   \mathrm{tr}\bigl( {\tilde{A}}^{\ast} {\tilde{A}} \bigr)
\nonumber \\ &
+ \tilde{K}_{P}^{(0)}
  \Bigl(
        \nabla_{\tilde{x}i} {\tilde{A}}^{ba\ast}
        \nabla_{\tilde{x}i} {\tilde{A}}^{ab}
     + \nabla_{\tilde{x}i} {\tilde{A}}^{ia\ast}
        \nabla_{\tilde{x}j} {\tilde{A}}^{aj}
     + \nabla_{\tilde{x}i} {\tilde{A}}^{ja\ast}
        \nabla_{\tilde{x}j} {\tilde{A}}^{ai}
  \Bigr)
\nonumber \\ &
+ \tilde{\beta}_{P}^{(0)}
   \Bigl(
        \mathrm{tr}\bigl( {\tilde{A}}^{\ast} {\tilde{A}} \bigr) \mathrm{tr}\bigl( {\tilde{A}}^{\ast} {\tilde{A}} \bigr)
      - \mathrm{tr}\bigl( {\tilde{A}}^{\ast} {\tilde{A}}^{\ast} {\tilde{A}} {\tilde{A}} \bigr)
   \Bigr)
\nonumber \\ &
+ \tilde{\gamma}_{P}^{(0)}
   \Bigl(
         - 3 \, \mathrm{tr}\bigl( {\tilde{A}} {\tilde{A}}^{\ast} \bigr) \, \mathrm{tr}\bigl( {\tilde{A}} {\tilde{A}} \bigr) \, \mathrm{tr}\bigl( {\tilde{A}}^{\ast} {\tilde{A}}^{\ast} \bigr)
        + 4 \, \mathrm{tr}\bigl( {\tilde{A}} {\tilde{A}}^{\ast} \bigr) \, \mathrm{tr}\bigl( {\tilde{A}} {\tilde{A}}^{\ast} \bigr) \, \mathrm{tr}\bigl( {\tilde{A}} {\tilde{A}}^{\ast} \bigr)
              \nonumber \\ & \hspace{3em} 
        + 6 \, \mathrm{tr}\bigl( {\tilde{A}}^{\ast} {\tilde{A}} \bigr) \, \mathrm{tr}\bigl( {\tilde{A}}^{\ast} {\tilde{A}}^{\ast} {\tilde{A}} {\tilde{A}} \bigr)
      + 12 \, \mathrm{tr}\bigl( {\tilde{A}}^{\ast} {\tilde{A}} \bigr) \, \mathrm{tr}\bigl( {\tilde{A}}^{\ast} {\tilde{A}} {\tilde{A}}^{\ast} {\tilde{A}} \bigr)
              \nonumber \\ & \hspace{3em} 
         - 6 \, \mathrm{tr}\bigl( {\tilde{A}}^{\ast} {\tilde{A}}^{\ast} \bigr) \, \mathrm{tr}\bigl( {\tilde{A}}^{\ast} {\tilde{A}} {\tilde{A}} {\tilde{A}} \bigr)
         - 6 \, \mathrm{tr}\bigl( {\tilde{A}} {\tilde{A}} \bigr) \, \mathrm{tr}\bigl( {\tilde{A}}^{\ast} {\tilde{A}}^{\ast} {\tilde{A}}^{\ast} {\tilde{A}} \bigr)
              \nonumber \\ & \hspace{3em} 
       - 12 \, \mathrm{tr}\bigl( {\tilde{A}}^{\ast} {\tilde{A}}^{\ast} {\tilde{A}}^{\ast} {\tilde{A}} {\tilde{A}} {\tilde{A}} \bigr)
      + 12 \, \mathrm{tr} \bigl( {\tilde{A}}^{\ast} {\tilde{A}}^{\ast} {\tilde{A}} {\tilde{A}} {\tilde{A}}^{\ast} {\tilde{A}} \bigr)
        + 8 \, \mathrm{tr}\bigl( {\tilde{A}}^{\ast} {\tilde{A}} {\tilde{A}}^{\ast} {\tilde{A}} {\tilde{A}}^{\ast} {\tilde{A}} \bigr)
   \Bigr)
\nonumber \\ &
 + \tilde{\delta}_{P}^{(0)}
\Bigl(
       \bigl( \mathrm{tr}\,\tilde{A}^{\ast 2} \bigr)^{2} \bigl( \mathrm{tr}\, \tilde{A}^{2} \bigr)^{2}
 + 2 \bigl( \mathrm{tr}\,\tilde{A}^{\ast 2} \bigr)^{2} \bigl( \mathrm{tr}\, \tilde{A}^{4} \bigr)
  - 8 \bigl( \mathrm{tr}\,\tilde{A}^{\ast 2} \bigr) \bigl( \mathrm{tr}\,\tilde{A}^{\ast}\tilde{A}\tilde{A}^{\ast}\tilde{A} \bigr) \bigl( \mathrm{tr}\,\tilde{A}^{2} \bigr)
  - 8 \bigl( \mathrm{tr}\,\tilde{A}^{\ast 2} \bigr) \bigl( \mathrm{tr}\,\tilde{A}^{\ast}\tilde{A} \bigr)^{2} \bigl( \mathrm{tr}\,\tilde{A}^{2} \bigr)
       \nonumber \\ & \hspace{3em} 
 - 32 \bigl( \mathrm{tr}\,\tilde{A}^{\ast 2} \bigr) \bigl( \mathrm{tr}\,\tilde{A}^{\ast}\tilde{A} \bigr) \bigl( \mathrm{tr}\,\tilde{A}^{\ast}\tilde{A}^{3} \bigr)
 - 32 \bigl( \mathrm{tr}\,\tilde{A}^{\ast 2} \bigr) \bigl( \mathrm{tr}\,\tilde{A}^{\ast}\tilde{A}\tilde{A}^{\ast}\tilde{A}^{3} \bigr)
 - 16 \bigl( \mathrm{tr}\,\tilde{A}^{\ast 2} \bigr) \bigl( \mathrm{tr}\,\tilde{A}^{\ast}\tilde{A}^{2}\tilde{A}^{\ast}\tilde{A}^{2} \bigr)
       \nonumber \\ & \hspace{3em} 
  + 2 \bigl( \mathrm{tr}\,\tilde{A}^{\ast 4} \bigr) \bigl( \mathrm{tr}\,\tilde{A}^{2} \bigr)^{2}
  + 4 \bigl( \mathrm{tr}\,\tilde{A}^{\ast 4} \bigr) \bigl( \mathrm{tr}\,\tilde{A}^{4} \bigr)
  - 32 \bigl( \mathrm{tr}\,\tilde{A}^{\ast 3}\tilde{A} \bigr) \bigl( \mathrm{tr}\,\tilde{A}^{\ast}\tilde{A} \bigr) \bigl( \mathrm{tr}\,\tilde{A}^{2} \bigr)
       \nonumber \\ & \hspace{3em} 
  - 64 \bigl( \mathrm{tr}\,\tilde{A}^{\ast 3}\tilde{A} \bigr) \bigl( \mathrm{tr}\,\tilde{A}^{\ast}\tilde{A}^{3} \bigr)
  - 32 \bigl( \mathrm{tr}\,\tilde{A}^{\ast 3}\tilde{A}\tilde{A}^{\ast}\tilde{A} \bigr) \bigl( \mathrm{tr}\,\tilde{A}^{2} \bigr)
  - 64 \bigl( \mathrm{tr}\,\tilde{A}^{\ast 3}\tilde{A}^{2}\tilde{A}^{\ast}\tilde{A}^{2} \bigr)
  - 64 \bigl( \mathrm{tr}\,\tilde{A}^{\ast 3}\tilde{A}^{3} \bigr) \bigl( \mathrm{tr}\,\tilde{A}^{\ast}\tilde{A} \bigr)
       \nonumber \\ & \hspace{3em} 
  - 64 \bigl( \mathrm{tr}\,\tilde{A}^{\ast 2}\tilde{A}\tilde{A}^{\ast 2}\tilde{A}^{3} \bigr)
  - 64 \bigl( \mathrm{tr}\,\tilde{A}^{\ast 2}\tilde{A}\tilde{A}^{\ast}\tilde{A}^{2} \bigr) \bigl( \mathrm{tr}\,\tilde{A}^{\ast}\tilde{A} \bigr)
 + 16 \bigl( \mathrm{tr}\,\tilde{A}^{\ast 2}\tilde{A}^{2} \bigr)^{2}
 + 32 \bigl( \mathrm{tr}\,\tilde{A}^{\ast 2}\tilde{A}^{2} \bigr) \bigl( \mathrm{tr}\,\tilde{A}^{\ast}\tilde{A} \bigr)^{2}
       \nonumber \\ & \hspace{3em} 
 + 32 \bigl( \mathrm{tr}\,\tilde{A}^{\ast 2}\tilde{A}^{2} \bigr) \bigl( \mathrm{tr}\,\tilde{A}^{\ast}\tilde{A}\tilde{A}^{\ast}\tilde{A} \bigr)
 + 64 \bigl( \mathrm{tr}\,\tilde{A}^{\ast 2}\tilde{A}^{2}\tilde{A}^{\ast 2}\tilde{A}^{2} \bigr)
  -16 \bigl( \mathrm{tr}\,\tilde{A}^{\ast 2}\tilde{A}\tilde{A}^{\ast 2}\tilde{A} \bigr) \bigl( \mathrm{tr}\,\tilde{A}^{2} \bigr)
   + 8 \bigl( \mathrm{tr}\,\tilde{A}^{\ast}\tilde{A} \bigr)^{4}
       \nonumber \\ & \hspace{3em} 
 + 48 \bigl( \mathrm{tr}\,\tilde{A}^{\ast}\tilde{A} \bigr)^{2} \bigl( \mathrm{tr}\,\tilde{A}^{\ast}\tilde{A}\tilde{A}^{\ast}\tilde{A} \bigr)
 +192 \bigl( \mathrm{tr}\,\tilde{A}^{\ast}\tilde{A} \bigr) \bigl( \mathrm{tr}\,\tilde{A}^{\ast}\tilde{A}\tilde{A}^{\ast 2}\tilde{A}^{2} \bigr)
 + 64 \bigl( \mathrm{tr}\,\tilde{A}^{\ast}\tilde{A} \bigr) \bigl( \mathrm{tr}\,\tilde{A}^{\ast}\tilde{A}\tilde{A}^{\ast}\tilde{A}\tilde{A}^{\ast}\tilde{A} \bigr)
       \nonumber \\ & \hspace{3em} 
  -128 \bigl( \mathrm{tr}\,\tilde{A}^{\ast}\tilde{A}\tilde{A}^{\ast 3}\tilde{A}^{3} \bigr)
 + 64 \bigl( \mathrm{tr}\,\tilde{A}^{\ast}\tilde{A}\tilde{A}^{\ast 2}\tilde{A}\tilde{A}^{\ast}\tilde{A}^{2} \bigr)
 + 24 \bigl( \mathrm{tr}\,\tilde{A}^{\ast}\tilde{A}\tilde{A}^{\ast}\tilde{A} \bigr)^{2}
 +128 \bigl( \mathrm{tr}\,\tilde{A}^{\ast}\tilde{A}\tilde{A}^{\ast}\tilde{A}\tilde{A}^{\ast 2}\tilde{A}^{2} \bigr)
       \nonumber \\ & \hspace{3em} 
 + 48 \bigl( \mathrm{tr}\,\tilde{A}^{\ast}\tilde{A}\tilde{A}^{\ast}\tilde{A}\tilde{A}^{\ast}\tilde{A}\tilde{A}^{\ast}\tilde{A} \bigr)
\Bigr)
   \nonumber \\ & 
+ \tilde{\beta}_{P}^{(2)}
      \vec{b}^{t} {\tilde{A}} {\tilde{A}}^{\ast} \vec{b}
+ \tilde{\beta}_{P}^{(4)}
   |\vec{b}|^{2}
   \vec{b}^{t} {\tilde{A}} {\tilde{A}}^{\ast} \vec{b},
   \nonumber \\ & 
+ \tilde{\gamma}_{P}^{(2)}
  \Bigl(
       - 2 \, |\vec{b}|^{2} \, \mathrm{tr}\bigl( {\tilde{A}} {\tilde{A}} \bigr) \, \mathrm{tr}\bigl( {\tilde{A}}^{\ast} {\tilde{A}}^{\ast} \bigr)
       - 4 \, |\vec{b}|^{2} \, \mathrm{tr}\bigl( {\tilde{A}} {\tilde{A}}^{\ast} \bigr) \, \mathrm{tr}\bigl( {\tilde{A}} {\tilde{A}}^{\ast} \bigr)
      + 4 \, |\vec{b}|^{2} \, \mathrm{tr}\bigl( {\tilde{A}} {\tilde{A}}^{\ast} {\tilde{A}} {\tilde{A}}^{\ast} \bigr)
      + 8 \, |\vec{b}|^{2} \, \mathrm{tr}\bigl( {\tilde{A}} {\tilde{A}} {\tilde{A}}^{\ast} {\tilde{A}}^{\ast} \bigr)
            \nonumber \\ & \hspace{2em} 
        + \vec{b}^{t} {\tilde{A}} {\tilde{A}} \vec{b} \, \mathrm{tr}\bigl( {\tilde{A}}^{\ast} {\tilde{A}}^{\ast} \bigr)
       - 8 \, \vec{b}^{t} {\tilde{A}} {\tilde{A}}^{\ast} \vec{b} \, \mathrm{tr}\bigl( {\tilde{A}} {\tilde{A}}^{\ast} \bigr)
         + \vec{b}^{t} {\tilde{A}}^{\ast} {\tilde{A}}^{\ast} \vec{b} \, \mathrm{tr}\bigl( {\tilde{A}} {\tilde{A}} \bigr)
      + 2 \, \vec{b}^{t} {\tilde{A}} {\tilde{A}}^{\ast} {\tilde{A}}^{\ast} {\tilde{A}} \vec{b}
            \nonumber \\ & \hspace{2em} 
      + 2 \, \vec{b}^{t} {\tilde{A}}^{\ast} {\tilde{A}} {\tilde{A}} {\tilde{A}}^{\ast} \vec{b}
       - 8 \, \vec{b}^{t} {\tilde{A}} {\tilde{A}}^{\ast} {\tilde{A}} {\tilde{A}}^{\ast} \vec{b}
       - 8 \, \vec{b}^{t} {\tilde{A}} {\tilde{A}} {\tilde{A}}^{\ast} {\tilde{A}}^{\ast} \vec{b}
  \Bigr),
\label{eq:free_energy_twice_dimensionless_P}
\end{align}
for the $^{3}P_{2}$ condensate, and
\begin{align}
    \tilde{f}_{SP}[\tilde{\sigma},\tilde{A}]
& =
   \tilde{\beta}_{SP}^{(0)}
   \Bigl(
        4 \, \tilde{\sigma}^{\ast} \tilde{\sigma} \, \dtr \bigl( \tilde{A}^{\ast} \tilde{A} \bigr)
         - \tilde{\sigma}^{2} \, \dtr \bigl( \tilde{A}^{\ast 2} \bigr)
         - \tilde{\sigma}^{\ast 2} \, \dtr \bigl( \tilde{A}^{2} \bigr)
   \Bigr),
\label{eq:free_energy_twice_dimensionless_SP}
\end{align}
for the coupling between the $^{1}S_{0}$ and $^{3}P_{2}$ condensates.
Notice $\nabla_{\tilde{x}i}=\partial/\partial\tilde{x}_{i}$ ($i=1,2,3$).
The GL coefficients can be calculated as
\begin{align}
   \tilde{\alpha}_{S}^{(0)} &= \frac{T-T_{Sc0}}{T_{Sc0}}, \quad 
   \tilde{K}_{S}^{(0)} = \frac{7\zeta(3)}{48\pi^{2}}, \quad 
   \tilde{\beta}_{S}^{(0)} = \frac{7\zeta(3)}{16\pi^{2}}, \quad 
   \tilde{\beta}_{S}^{(2)} = \frac{7\zeta(3)}{16\pi^{2}}, \quad \nonumber \\ 
   \tilde{\alpha}_{P}^{(0)} &= \frac{1}{3} \frac{T-T_{Pc0}}{T_{Pc0}}, \quad 
   \tilde{K}_{P}^{(0)} = \frac{7 \, \zeta(3)}{240\pi^{2}}, \quad 
   \tilde{\beta}_{P}^{(0)} = \frac{7\,\zeta(3)}{60\pi^{2}}, \quad 
   \tilde{\gamma}_{P}^{(0)} = - \frac{31\,\zeta(5)}{13440\pi^{4}}, \quad 
   \tilde{\delta}_{P}^{(0)} = \frac{127\,\zeta(7)}{387072\pi^{6}}, \nonumber \\
   \tilde{\beta}_{P}^{(2)} &= \frac{7\,\zeta(3)}{48\pi^{2}}, \quad 
   \tilde{\beta}_{P}^{(4)} = - \frac{31\,\zeta(5)}{768\pi^{4}}, \quad 
   \tilde{\gamma}_{P}^{(2)} = \frac{31\,\zeta(5)}{3840\pi^{4}}, \nonumber \\ 
   \tilde{\beta}_{SP}^{(0)} &= \frac{7\zeta(3)}{48\pi^{2}}.
\label{eq:free_energy_twice_dimensionless_coefficients}
\end{align}
To derive the above equations, we have adopted the following approximations: $\ln(T/T_{c0})\approx(T-T_{c0})/T_{c0}$ with $T$ being close to the critical temperature $T_{c0}$.
We assume that the critical temperatures $T_{Sc0}$ and $T_{Pc0}$ for the neutron $^{1}S_{0}$ superfluid ($\tilde{\sigma}$) and the neutron $^{3}P_{2}$ superfluid ($\tilde{A}$), respectively, coincide: $T_{c0}=T_{Sc0}=T_{Pc0}$.
We define $t=T/T_{c0}$ as a dimensionless quantity.
This condition guarantees that the GL expansion is applicable because both  $\tilde{\sigma}$ and $\tilde{A}$ are small quantities. 
This situation can be realized in the neutron stars by the following reason.
As the number density of the neutron matter increases the interaction strength in the $^{1}S_{0}$ channel decreases, and the attraction at the lower density turns to a repulsion at the higher density. On the other hand, the interaction strength in the $^{3}P_{2}$ channel increases from the lower density to the higher density. Therefore, there should exist the baryon number density where the superfluid transitions in the $^{1}S_{0}$ and $^{3}P_{2}$ channels occur at the same critical temperature.

We emphasize that the coupling term, Eq.~\eqref{eq:free_energy_twice_dimensionless_SP}, is a new term found for the first time in the present paper.
This term can describe the coexistence of the $^{1}S_{0}$ and $^{3}P_{2}$ superfluids.
In the next section, we will find that this term changes the properties of the $^{3}P_{2}$ condensate due to the existence of the $^{1}S_{0}$ condensate.

We leave some comments for each term in the GL free energy~\eqref{eq:free_energy_twice_dimensionless}.
In Eq.~\eqref{eq:free_energy_twice_dimensionless_S} for the $^{1}S_{0}$ condensate, it is apparent that the second-order term with $\tilde{\alpha}_{S}^{(0)}$ induces the nonzero condensate with the symmetry breaking for $t<1$, the fourth-order term with $\tilde{\beta}_{S}^{(0)}$ supports the stability of the ground state, and the $\tilde{\beta}_{S}^{(2)}$ term plays the role to recover the broken symmetry by the magnetic field.

On the other hand, the situation in the $^{3}P_{2}$ condensate is more complex.
The second-order term with $\tilde{\alpha}_{P}^{(0)}$ induces the nonzero condensate for $t<1$ and the fourth-order term with $\tilde{\beta}_{P}^{(0)}$ supports the stability of the ground state.
However, the ground state is not uniquely determined at the fourth order, and there remains the continuous degeneracy among the UN, D$_{2}$-BN, and D$_{4}$-BN phases.
This degeneracy can be resolved by the sixth-order term with $\tilde{\gamma}_{P}^{(0)}$, which, however, gives only a local minimum for the ground state, yielding the instability for large values of the condensate.
The global stability of the ground state is provided by the eighth-order term with $\tilde{\delta}_{P}^{(0)}$~\cite{Yasui:2019unp}.
In Ref.~\cite{Yasui:2019unp}, it was shown that the GL expansion up to the eighth order describes the first-order phase transition known in the analysis of the BdG equation~\cite{Mizushima:2016fbn}. It can further capture the CEP at a meeting point of the first-order and  second-order phase transition lines.
The critical exponents at the CEP were analyzed by the GL equation as well as the BdG equation~\cite{Mizushima:2019spl}.
The  $\tilde{\beta}_{P}^{(2)}$, $\tilde{\beta}_{P}^{(4)}$, and $\tilde{\gamma}_{P}^{(2)}$ terms describe responses to the magnetic field.
The $\tilde{\beta}_{P}^{(2)}$ term is the leading-order term and the higher-order $\tilde{\beta}_{P}^{(4)}$ and $\tilde{\gamma}_{P}^{(2)}$ terms 
were calculated to investigate the change of the phase diagram for strong magnetic fields, relevant for magnetars~\cite{Yasui:2018tcr}; in this case, the phase boundary is changed by $\approx$ 10\% at most.

In the present paper, we consider that the critical temperature of the $^{1}S_{0}$ superfluids is tuned in a way independent from the $^{3}P_{2}$ superfluids and assume that they coincide with each other at a certain baryon number density. We comment, however, that there is a possibility that those critical temperatures are different from each other when the coupling between the $^{1}S_{0}$ pairing and the $^{3}P_{2}$ pairing is taken into account microscopically. This study is left for future work.

\section{Numerical results}
\label{sec:numerical_results}
\begin{figure}[t]
\begin{center}
\vspace{0cm}
\includegraphics[scale=0.3]{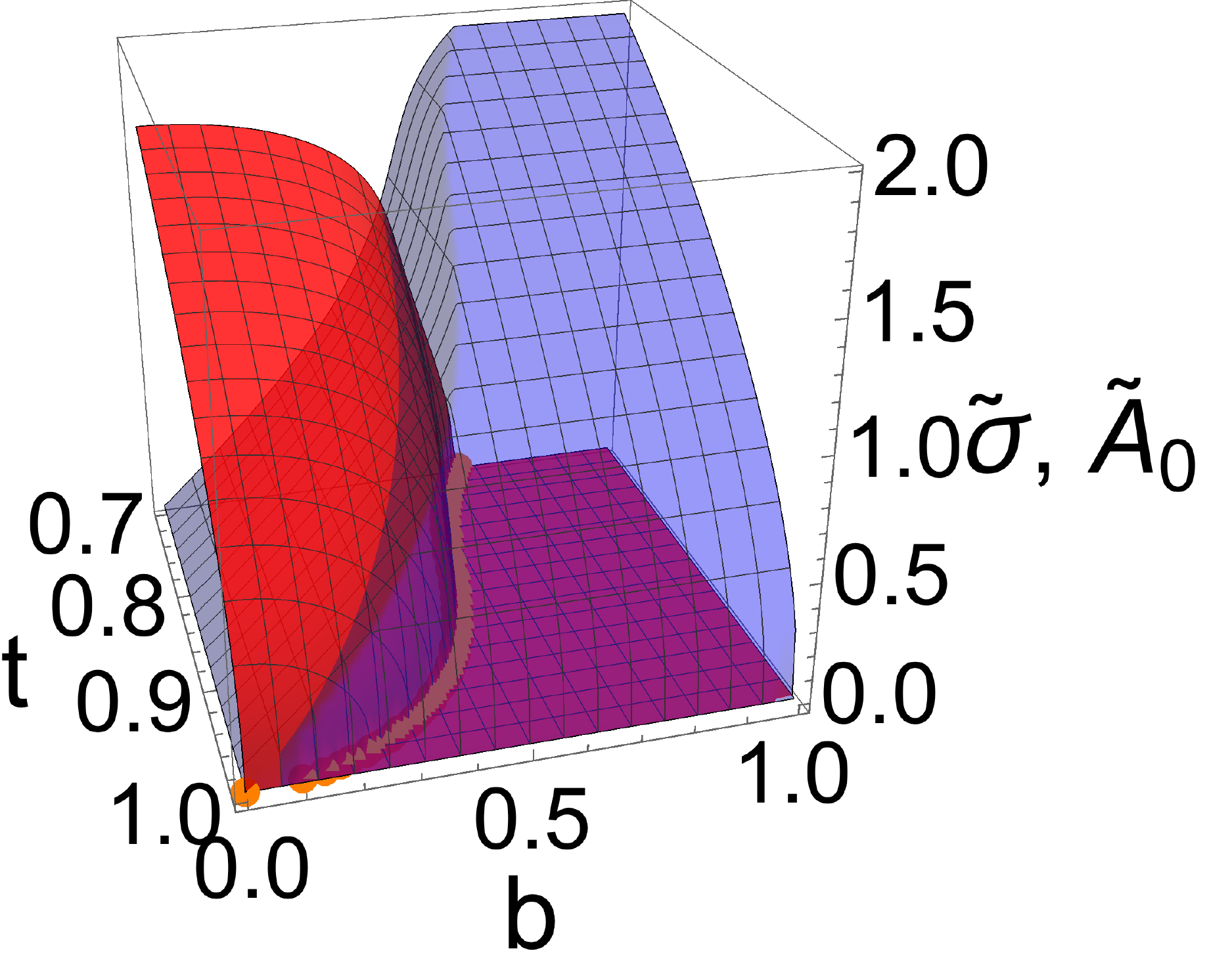}
\vspace{0em}
\caption{Plots of $\tilde{\sigma}$ (red surface) and $\tilde{A}_{0}$ (blue surface) as functions of $t$ and $b$. We also show the phase boundary (the second-order phase transition) of the $^{1}S_{0}$ superfluid in the presence of the $^{3}P_{2}$ superfluid.}
\label{fig:sigma_tau0_3d_t_b}
\end{center}
\end{figure}

\begin{figure}[t]
\begin{center}
\vspace{0cm}
\includegraphics[scale=0.25]{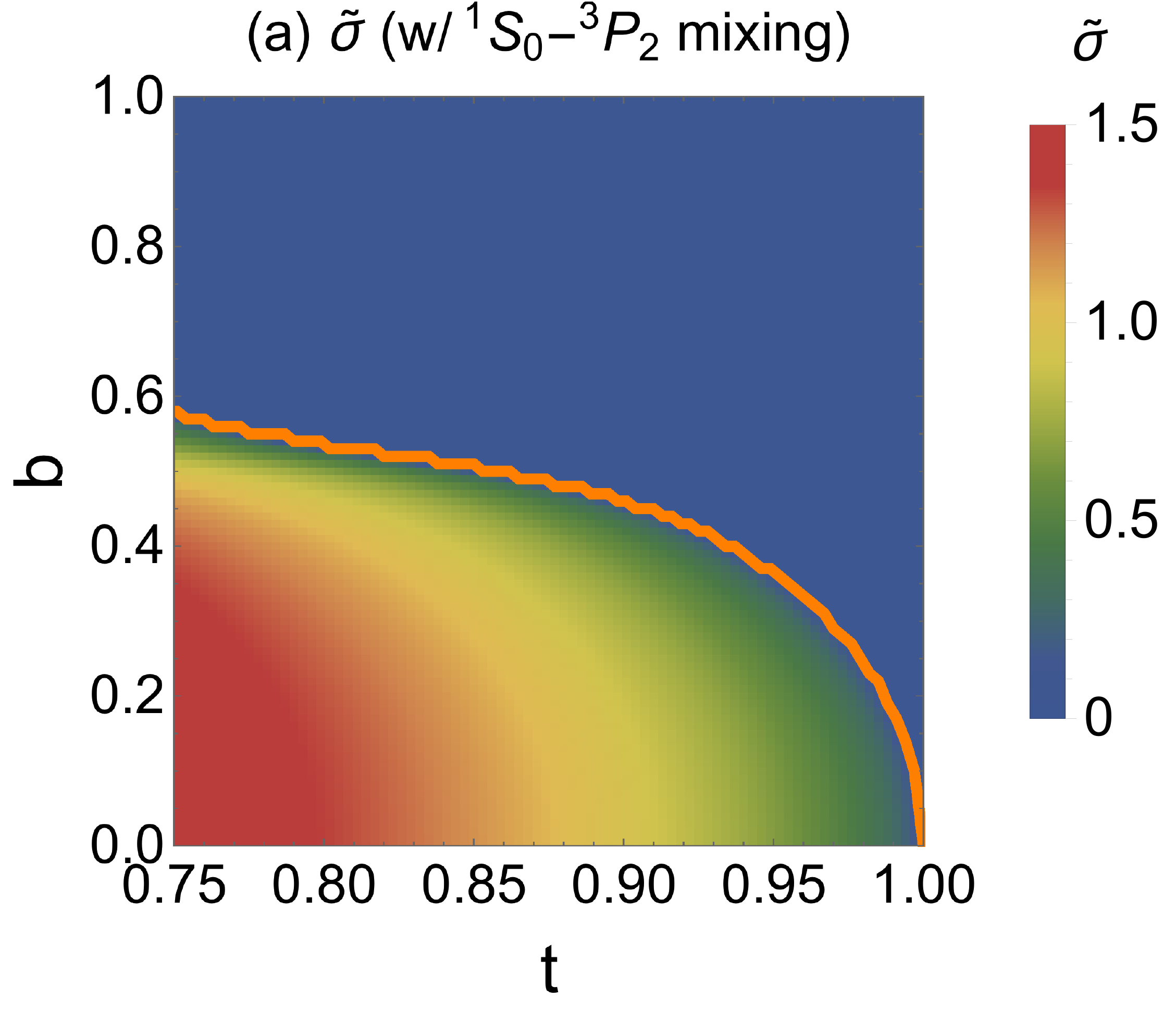}
\includegraphics[scale=0.25]{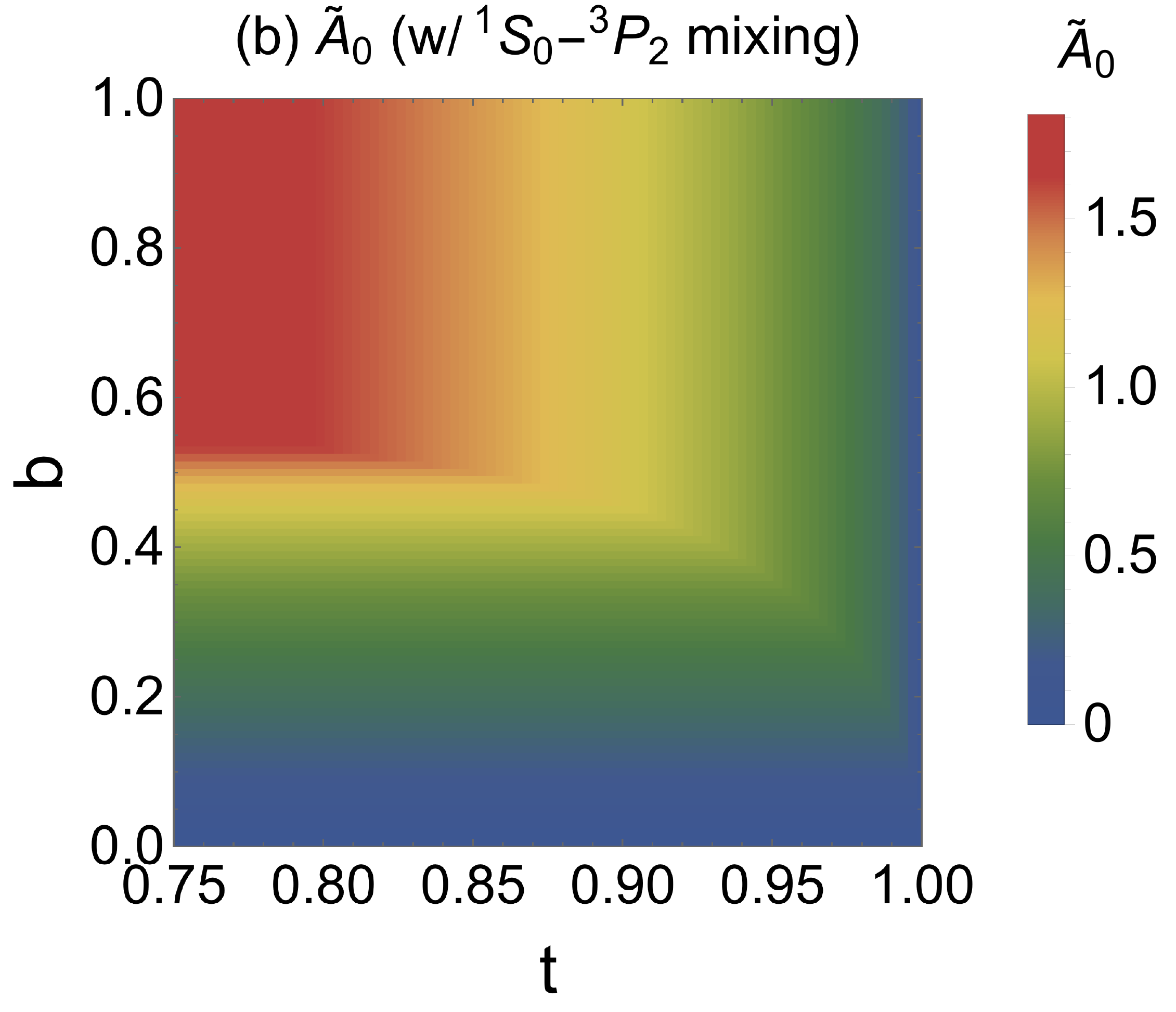}
\includegraphics[scale=0.25]{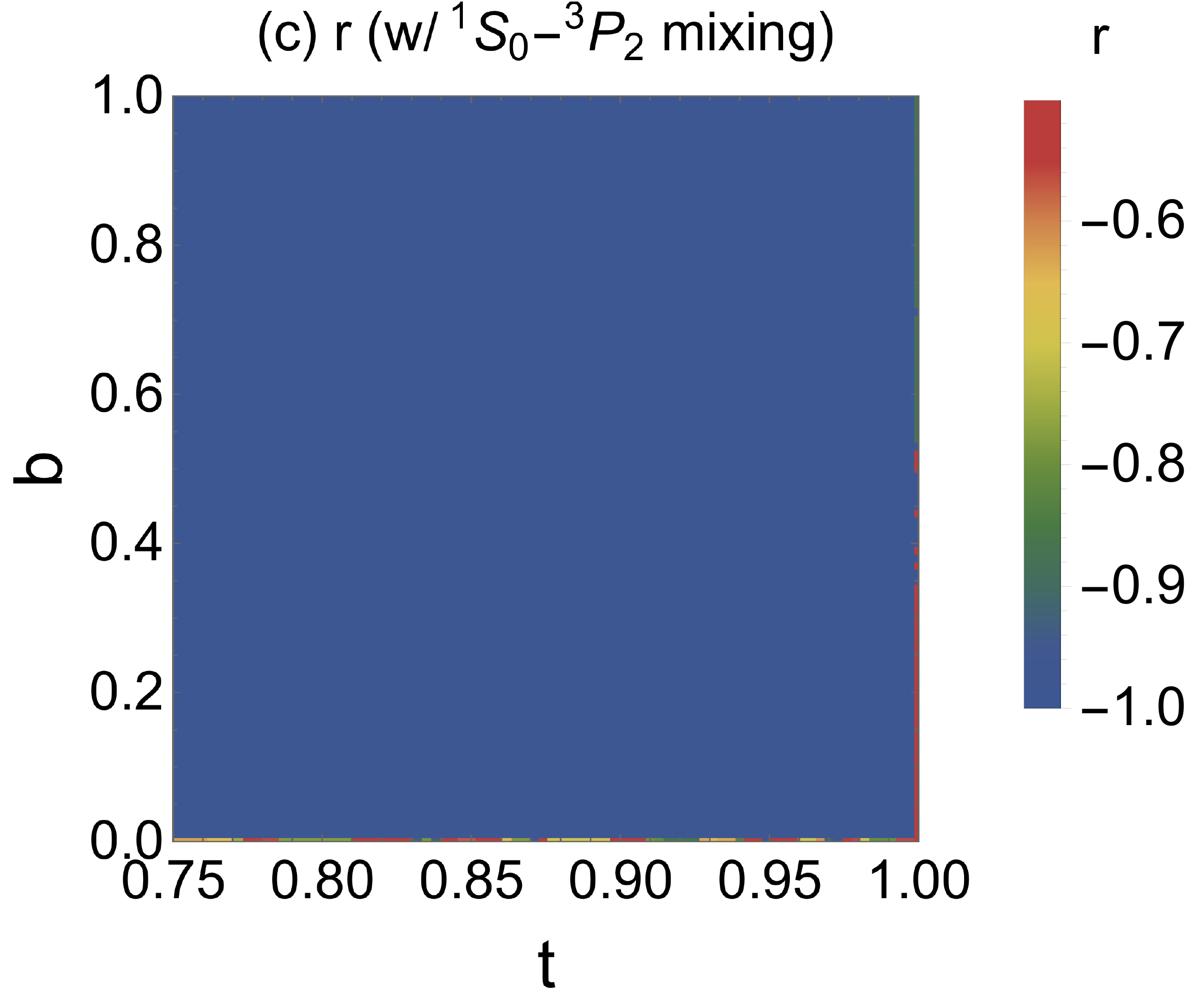}
\\
\includegraphics[scale=0.25]{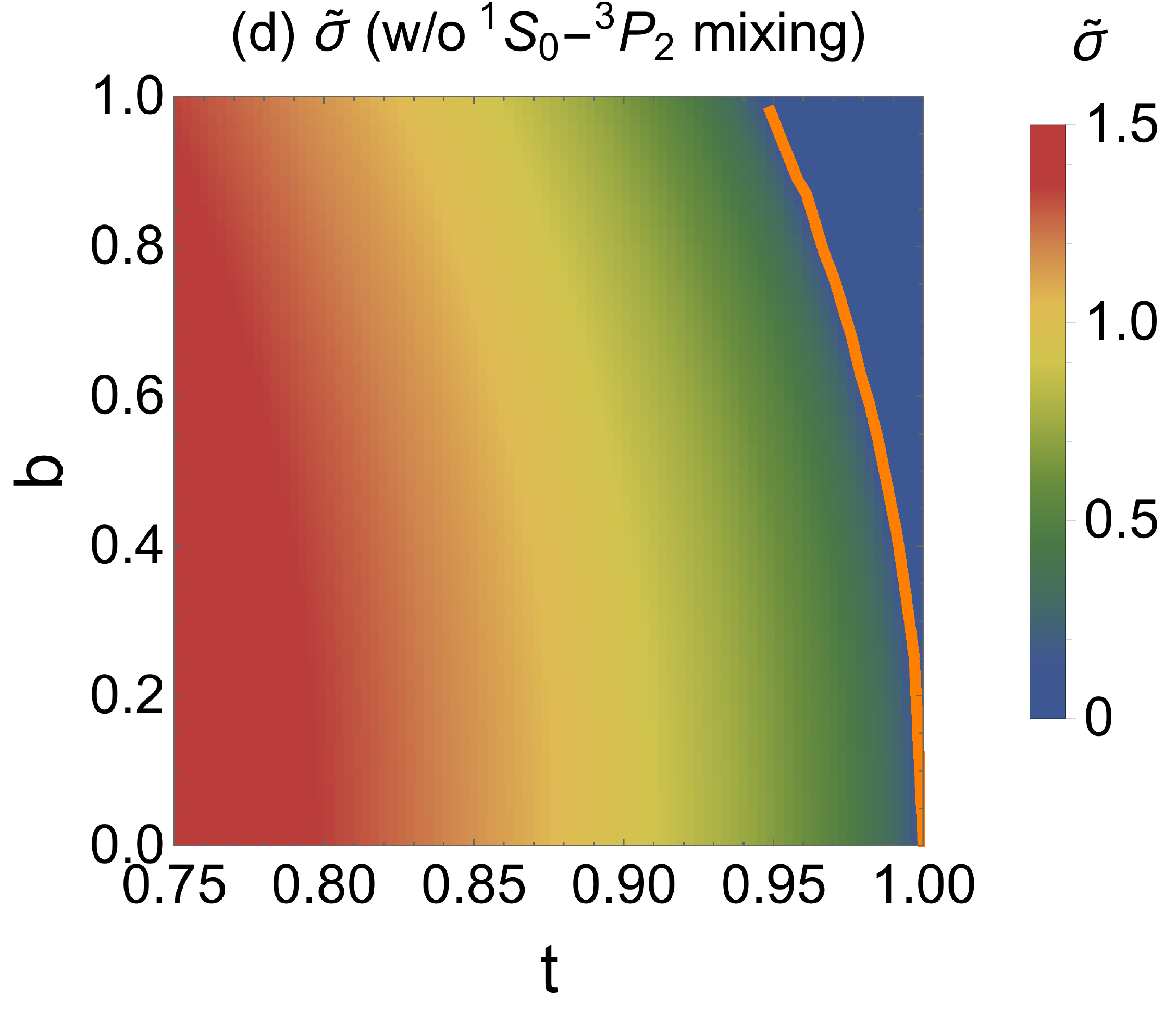}
\includegraphics[scale=0.25]{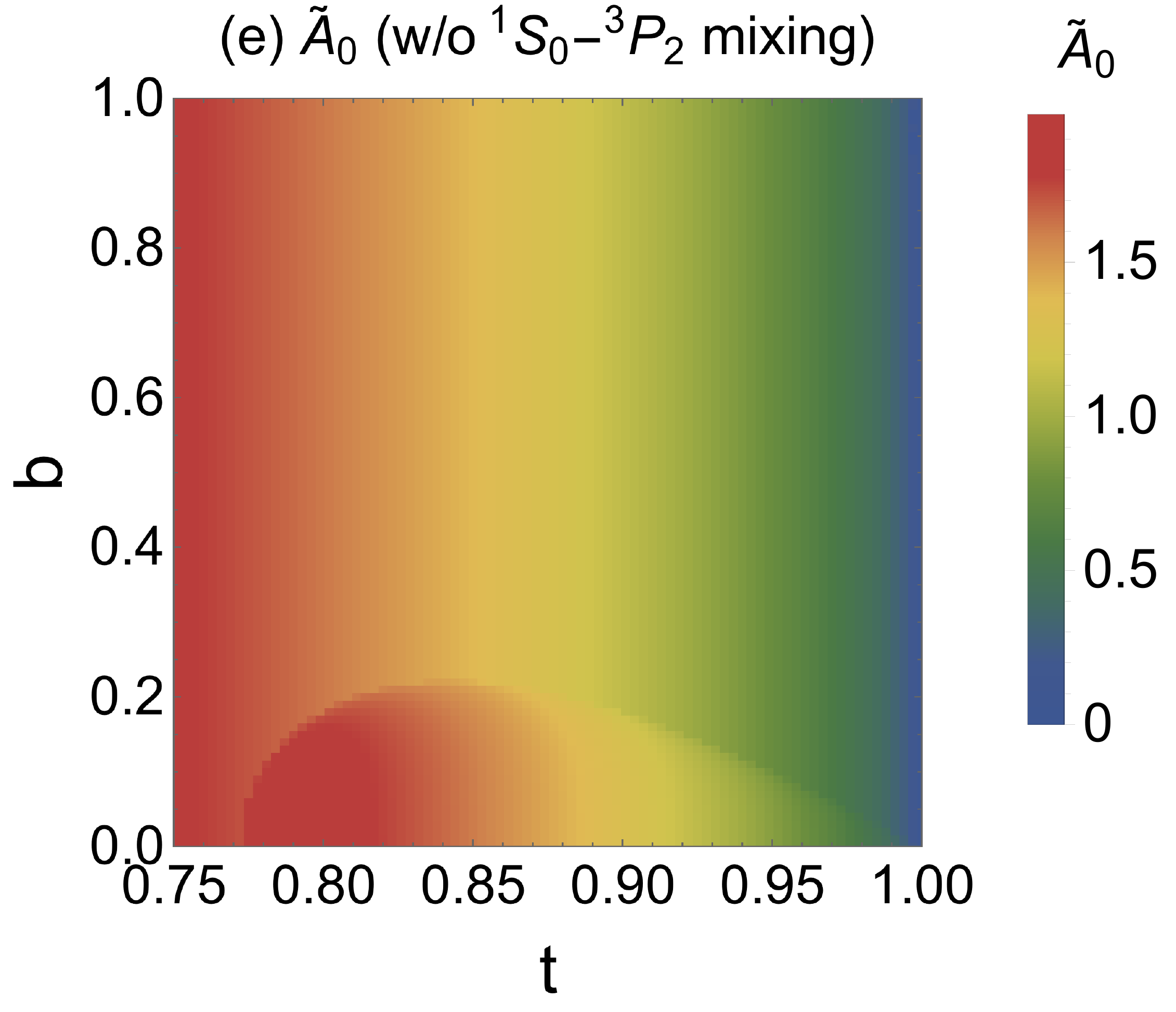}
\includegraphics[scale=0.25]{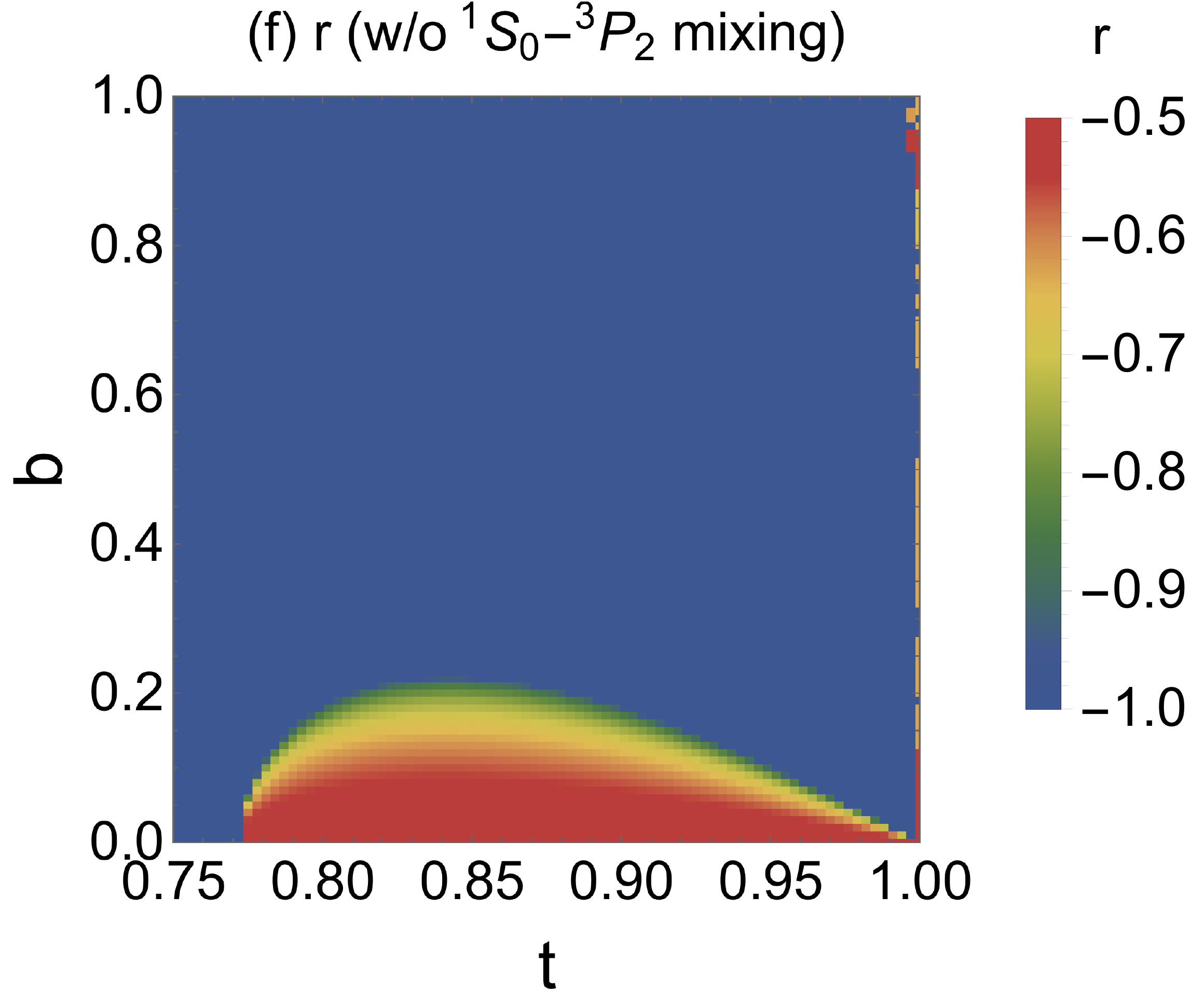}
\vspace{0em}
\caption{Phase diagrams of $^{1}S_{0}$ superfluid ($\tilde{\sigma}$) and $^{3}P_{2}$ superfluid ($\tilde{A}_{0}$ and $r$) in neutron gas with respect to (dimensionless) temperature $t$ and magnetic field $b$. The first row is for the full coupling in the $^{1}S_{0}$-$^{3}P_{2}$ coupling ($\tilde{\beta}_{SP}^{(0)}\neq0$) and the second row is for the case that the $^{1}S_{0}$-$^{3}P_{2}$ coupling is turned off ($\tilde{\beta}_{SP}^{(0)}=0$). 
We also show the phase boundary (the second-order phase transition) of the $^{1}S_{0}$ superfluid in the presence of the $^{3}P_{2}$ superfluid in the panels (a) and (d).}
\label{fig:phase_diagram_wSP_t_b}
\end{center}
\end{figure}

In this section, based on the GL free energy \eqref{eq:GL_free_energy}, 
we analyze the phase diagram for 
 the $^{1}S_{0}$ superfluid $\tilde{\sigma}$ and 
 the $^{3}P_{2}$ superfluid $\tilde{A}$.
Without loss of generality, we can express $\tilde{A}$ in a diagonal form
\begin{align}
   \tilde{A}
=
\tilde{A}_{0}
\left(
\begin{array}{ccc}
 r & 0 & 0 \\
 0 & -1-r & 0 \\
 0 & 0 & 1
\end{array}
\right),
\label{eq:parametrization_tau}
\end{align}
by applying the appropriate SO(3) transformation to the original $\tilde{A}$.
Here $\tilde{A}_{0}$ ($\tilde{A}_{0}\ge0$) is the amplitude and a real parameter $r$ ($-1\le r \le -1/2$) characterizing the ground state.
According to the value of $r$, the ground state has the following symmetries: the O(2) symmetry for $r=-1/2$ (the UN phase), the D$_{2}$ symmetry for $-1<r<-1/2$ (the D$_{2}$-BN phase), and the D$_{4}$ symmetry for $r=-1$ (the D$_{4}$-BN phase).
The values of $\tilde{\sigma}$, $\tilde{A}_{0}$ and $r$ are determined to minimize
the GL free energy \eqref{eq:GL_free_energy}.
In Figs.~\ref{fig:sigma_tau0_3d_t_b} and \ref{fig:phase_diagram_wSP_t_b} (a), (b), and (c),
we show the calculated order parameters of
the $^{1}S_{0}$ and $^{3}P_{2}$ superfluids on the plane spanned by the dimensionless temperature $t$ and the magnetic field $b$ [see Eq.~\eqref{eq:dimensionless_quantities}].
Here Fig.~\ref{fig:sigma_tau0_3d_t_b} is the three-dimensional plots of Fig.~\ref{fig:phase_diagram_wSP_t_b} (a) and (b).
We notice that the temperature regions for $t>1$ are normal phase and those of $t<1$ are the superfluid phases.

\begin{figure}[t]
\begin{center}
\vspace{0cm}
\includegraphics[scale=0.3]{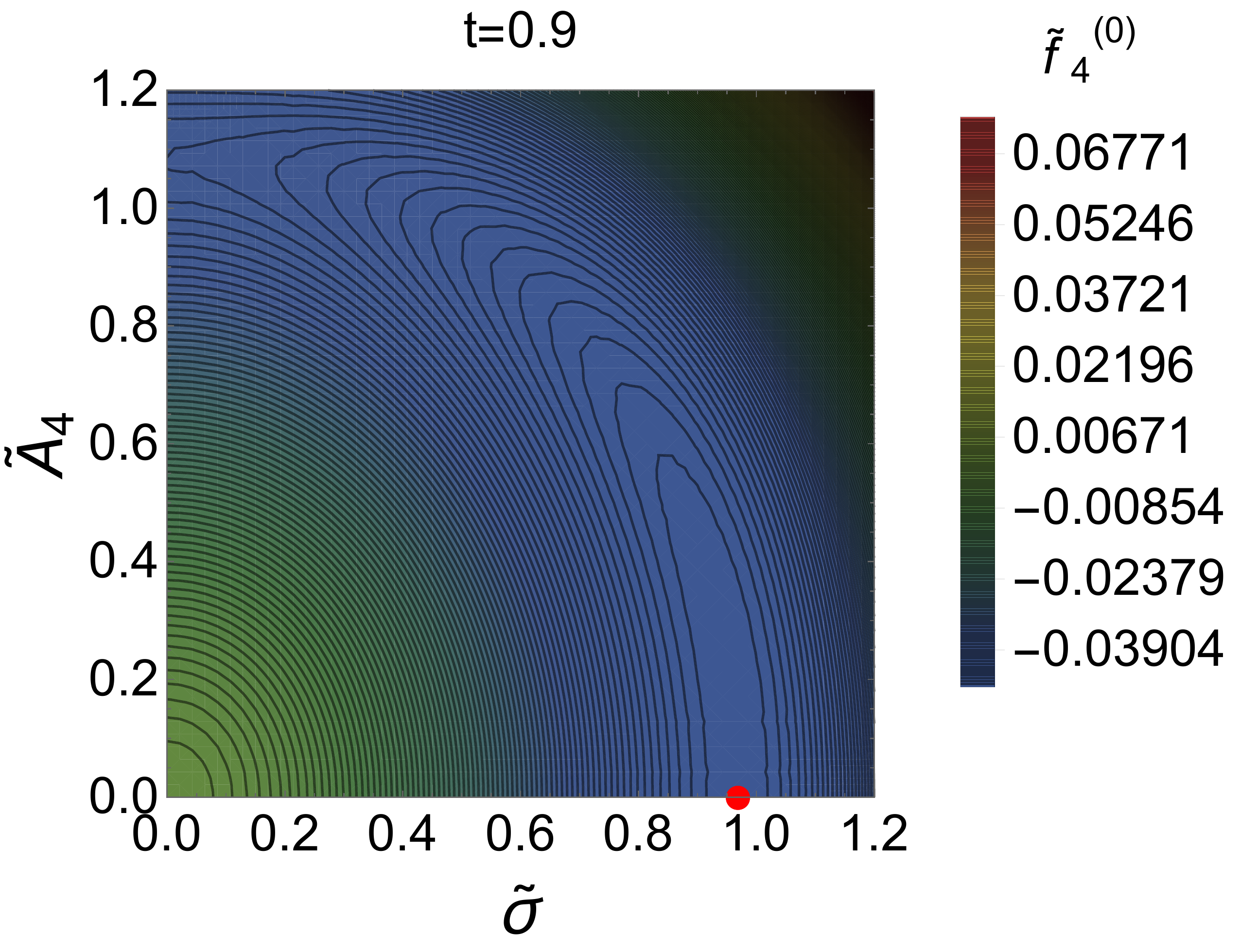}
\vspace{0em}
\caption{A plot of the dimensionless GL free energy $\tilde{f}_{4}^{(0)}$ up to the fourth order in the condensates $\tilde{\sigma}$  and $\tilde{A}_{4}$ at $t=0.9$. The red blob indicates the global minimum at $\tilde{\sigma} = \sqrt{{8\pi^{2}(1-t)}/({7\zeta(3)}})$ and $\tilde{A}_{4} = 0$.}
\label{fig:f4_sigma_A4_t090}
\end{center}
\end{figure}

In the absence of the magnetic field, only the $^{1}S_{0}$ superfluid is realized
 and the $^{3}P_{2}$ superfluid is excluded ($\tilde{A}_0=0$).
This can be analytically proved as follows.
We consider the dimensionless GL free energy up to the fourth-order term given by
\begin{align}
   \tilde{f}_{4}^{(0)}[\tilde{\sigma},\tilde{A}]
=
   \frac{t-1}{3}
   (3\tilde{\sigma}^{2}+2\tilde{A}_{4}^{2})
+ \frac{7\zeta(3)}{240\pi^{2}}
   (15\tilde{\sigma}^{4}+20\tilde{\sigma}^{2}\tilde{A}_{4}^{2}+8\tilde{A}_{4}^{4}),
\label{eq:free_energy_twice_dimensionless_4_tau0_r}
\end{align}
where $\tilde{A}_{4}^{2} = (1+r+r^{2}) \tilde{A}_{0}^{2}$.
It is sufficient for us to neglect the higher-order terms, because the existence or nonexistence ($\tilde{A}_{0}=0$ or $\tilde{A}_{0}\neq0$, i.e., $\tilde{A}_{4}=0$ or $\tilde{A}_{4}\neq0$) of the $^{3}P_{2}$ superfluid is determined by the fourth-order expansion.\footnote{Notice $1+r+r^{2} \neq 0$ for any $r$.
The role of the sixth-order term or the higher-order terms is to determine finely the internal symmetries for $\tilde{A}_{0} \neq 0$, i.e., the UN, D$_{2}$-BN, and D$_{4}$-BN phases.}
From the stationary condition of Eq.~\eqref{eq:free_energy_twice_dimensionless_4_tau0_r} with respect to $\tilde{\sigma}$ and $\tilde{A}_{4}$, 
we find that the solution $\tilde{\sigma} = \sqrt{{8\pi^{2}(1-t)}/({7\zeta(3)}})$ and $\tilde{A}_{4} = 0$ gives the globally minimum energy.
To be clear, we show the plot of $\tilde{f}_{4}^{(0)}[\tilde{\sigma},\tilde{A}]$ as a function of $\tilde{\sigma}$ and $\tilde{A}_{4}$ in Fig.~\ref{fig:f4_sigma_A4_t090} in which the global
 minimum is denoted by the red blob corresponding to the obtained solution.
Therefore, we conclude that only $^{1}S_{0}$ superfluid survives at zero magnetic field and the $^{3}P_{2}$ superfluid is completely suppressed by the coupling term, and hence there is no coexistence of the two phases for zero magnetic field.

In the presence of the magnetic fields, however, the situation of the (non)coexistence of the two superfluid phases changes qualitatively.
As for the $^{1}S_{0}$ superfluid, the value of $\tilde{\sigma}$ tends to decrease as the magnetic field becomes stronger due to the pair breaking associated with the Zeeman effects, and it eventually vanishes at a critical value of $b$.
Contrary to this, as for the $^{3}P_{2}$ superfluid, the value of $\tilde{A}_{0}$ is much suppressed in the weak magnetic-field regime, and it becomes enhanced as the magnetic field increases.
The internal order parameter ($r$) indicates that the D$_{4}$-BN phase is always stable for the $^{3}P_{2}$ superfluid.
In this way, we find that the $^{1}S_{0}$ superfluid and the $^{3}P_{2}$ superfluid with the D$_{4}$-BN phase coexist at the intermediate strengths of magnetic fields.
It is reasonable that the $^{1}S_{0}$ superfluid disappears at the strong magnetic fields, while the $^{3}P_{2}$ superfluid survives, 
because the spin-antiparallel pairing of two neutrons, $\uparrow\downarrow$, in the $^{1}S_{0}$ superfluid is destroyed by the strong magnetic fields, while the spin-parallel pairing of two neutrons, $\uparrow\uparrow$ or $\downarrow\downarrow$, in the $^{3}P_{2}$ superfluid with the D$_{4}$-BN phase remains to be unchanged.
Thus, the D$_{4}$-BN phase ($r=-1$) is favored in the $^{3}P_{2}$ superfluid in the presence of a magnetic field no matter how small it is. 

In the GL free energy \eqref{eq:GL_free_energy}, a possible coexistence of the $^{1}S_{0}$ and $^{3}P_{2}$ superfluids is determined by the coupling term in Eq.~\eqref{eq:free_energy_twice_dimensionless_SP}.
In order to illustrate the coupling effect,
we show the phase diagram of the $^{1}S_{0}$ and the $^{3}P_{2}$ superfluids by turning off the coupling between the two superfluids 
($\tilde{\beta}_{SP}^{(0)}=0$) in Fig.~\ref{fig:phase_diagram_wSP_t_b} (d), (e), and (f).
Notice that the results in Fig.~\ref{fig:phase_diagram_wSP_t_b} (e) and (f) were obtained in the previous work~\cite{Yasui:2019unp,Mizushima:2019spl}.
We find that the regions of the $^{1}S_{0}$ phase are different for Fig.~\ref{fig:phase_diagram_wSP_t_b} (a) and (d): The region of the $^{1}S_{0}$ phase is suppressed by the presence of the $^{3}P_{2}$ superfluid through the coupling term.
As for the $^{3}P_{2}$ phase, we notice that there exists not only the D$_{4}$-BN phase but also the UN  and D$_{2}$-BN phases when 
the coupling term is switched off, while there is only the D$_{4}$-BN phase once the coupling term is taken into account [see Figs.~\ref{fig:phase_diagram_wSP_t_b} (c) and (f)].
It is important that, as a result of the coupling effect, the D$_{4}$-BN phase exists even at weak magnetic fields.
This contrasts sharply with the case 
when the coupling term is switched off,
in which the D$_{4}$-BN phase exists only with the strong magnetic fields. 
This result indicates a possibility that the D$_{4}$-BN phase is realized not only in magnetars with strong magnetic fields but also in ordinary neutron stars with weak magnetic fields. 
It brings us more chances to have rich topological phenomena in the D$_{4}$-BN phase such as 
half-quantized non-Abelian vortices~\cite{Masuda:2016vak}.

\section{Conclusion and outlook}
\label{sec:conclusion}

We have discussed the coexistence of the $^{1}S_{0}$ and $^{3}P_{2}$ superfluids in neutron stars.
Starting from the interaction between two neutrons, we have adopted the weak-coupling limit and obtained the GL free energy with the  coupling term between the two superfluids.
We have analyzed the phase diagram and shown that the $^{1}S_{0}$ superfluid completely excludes the $^{3}P_{2}$ superfluid at zero magnetic field and both the superfluids can coexist at weak magnetic fields, and the $^{1}S_{0}$ superfluid is expelled by the $^{3}P_{2}$ superfluid at strong magnetic fields.
Remarkably it has been shown that the region of the D$_{4}$-BN phase is extended to the whole range of nonzero magnetic fields because of the coupling between the two superfluids. This result is in contrast to the case without the $^1S_0$ superfluid, where the D$_{4}$-BN phase is stable only in the strong magnetic-field region.
Our result indicates that the D$_{4}$-BN phase is realized not only in magnetars with strong magnetic fields but also in normal neutron stars with weak magnetic fields, indicating the importance of studies of various topological phenomena in the D$_{4}$-BN phase.

In the present paper, we have concentrated on the bosonic excitations within the GL equation.
More fundamentally, we can solve the BdG equation self-consistently to obtain the phases of the $^{3}P_{2}$ superfluid~\cite{Mizushima:2016fbn,Masaki:2019rsz}. 
It is known that, according to the general classifications,
the nematic phase in the neutron $^{3}P_{2}$ superfluids is a class-DIII topological superconductor in the periodic table, inducing Majorana fermions on the edge of the superfluids~\cite{Mizushima:2016fbn}.
The cyclic and ferromagnetic phases 
are nonunitary states, in which the time-reversal symmetry is broken, and they serve to host Weyl fermions in the bulk~\cite{Mizushima:2016fbn,Mizushima:2017pma}.
It will be an interesting question how such topological properties 
are modified in the coexistence of the $^{1}S_{0}$ and $^{3}P_{2}$ superfluids.
Quantized vortices in $^{3}P_{2}$ superfluids are also interesting as they were extensively studied in Refs.~\cite{Muzikar:1980as,Sauls:1982ie,Fujita1972,Richardson:1972xn,Masuda:2015jka,Chatterjee:2016gpm,Masuda:2016vak,Masaki:2019rsz}. 
It should be important to study how the coexistence of the $^{1}S_{0}$ and $^{3}P_{2}$ superfluids affects the properties of quantized vortices;
since two condensates coexist, 
vortices will become fractionally quantized
and vortices in different condensates 
weakly repel each other, 
as the case of miscible two-component BECs 
\cite{Eto:2011wp,Kasamatsu:2015cia}. 
It is also known that the vortices may terminate on a domain wall forming a so-called D-brane soliton, which exists in two-component BEC~\cite{Kasamatsu:2010aq, Nitta:2012hy, Kasamatsu:2013lda, Kasamatsu:2013qia} and supersymmetric field theory~\cite{Gauntlett:2000de, Isozumi:2004vg}, where the end point of the vortex is called a boojum. 
There is another possibility that a domain wall may terminate on a vortex; a vortex may be attached by a domain wall, as axion strings.
Those are also interesting objects to be explored in the coexistence phase of the $^{1}S_{0}$ and $^{3}P_{2}$ superfluids.

Because topological objects can also exist in quark matter, a connection between the hadronic phase and the quark phase should be studied in detail.
Then, it is an interesting question how vortices in the hadronic phase, such as those in the coexistence phase of the $^{1}S_{0}$ and $^{3}P_{2}$ superfluids, interact with other topological defects in the quark phase~\cite{Balachandran:2005ev,Nakano:2007dr,Eto:2009kg,Eto:2009bh,Eto:2009tr}.
In a rotating neutron star, non-Abelian quantum vortices (color magnetic flux tubes) are created  in the quark matter~\cite{Balachandran:2005ev,Nakano:2007dr,Eto:2009kg,Eto:2009bh,Eto:2009tr} (see also Ref.~\cite{Eto:2013hoa} as a review),
and they may lead to the presence or absence of boojums as defects at end points (or junction points) of these vortices at the interface (or the crossover region)~\cite{Cipriani:2012hr,Alford:2018mqj,Chatterjee:2018nxe,Chatterjee:2019tbz,Cherman:2018jir}.
The interactions between the vortices and boojums may influence the dynamics inside neutron stars.
Those problems are left for future studies.

\section*{Acknowledgment}
We thank Michikazu Kobayashi for useful comments.
This work is supported by the Ministry of Education, Culture, Sports, Science, and Technology (MEXT)-supported program for the Strategic Research Foundation at Private Universities ``Topological Science" (Grant No. S1511006). 
This work is also supported in part by 
a Japan Society for the Promotion of Science Grant-in-Aid for Scientific Research [KAKENHI Grants No.~17K05435 (S.Y.), 
No.~16H03984 (M.N.), and No.~18H01217 (M.N.)], 
and also by a MEXT KAKENHI Grant-in-Aid for Scientific Research on Innovative Areas ``Topological Materials Science'' Grant No.~15H05855 (M.N.).
\bibliography{neutronstar}

\end{document}